\documentstyle[twocolumn,prl,float,aps,epsfig,rotating]{revtex}

\begin{document}
\draft

\title{Measurement of the resonant $d \mu t$ molecular formation rate in solid 
HD}

\author{%
\mbox{T.A.~Porcelli$^{1,}$}\cite{newadd}, 
\mbox{A.~Adamczak$^2$,}
\mbox{J.M.~Bailey$^3$,}
\mbox{G.A.~Beer$^1$,} 
\mbox{J.L.~Douglas$^4$,}
\mbox{M.P.~Faifman$^5$,}
\mbox{M.C.~Fujiwara$^{4,}$}\cite{newadd2},
\mbox{T.M.~Huber$^6$,}
\mbox{P.~Kammel$^7$,}
\mbox{S.K.~Kim$^8$,}
\mbox{P.E.~Knowles$^9$,}
\mbox{A.R.~Kunselman$^{10}$,}
\mbox{M.~Maier$^1$,   } 
\mbox{V.E.~Markushin$^{11}$,}
\mbox{G.M.~Marshall$^{12}$,   }
\mbox{G.R.~Mason$^1$,} 
\mbox{F. Mulhauser$^{9}$,   }
\mbox{A.~Olin$^{1,12}$,} 
\mbox{C.~Petitjean$^{11}$,}
\mbox{J.~Zmeskal$^{13}$}
\\
(TRIUMF Muonic Hydrogen  Collaboration)
\\}

\address{%  
\mbox{$^1$ Department of Physics and Astronomy, University of Victoria, 
Victoria, BC, Canada V8W 3P6} 
\mbox{$^2$ Institute of Nuclear Physics, PL-31-342 Cracow, Poland} \\ 
\mbox{$^3$  Chester Technology, Chester, UK} \\ 
\mbox{$^4$  Department of Physics and Astronomy, University of British Columbia,
Vancouver, BC, Canada V6T 1Z1}
\mbox{$^5$ Russian Research Center, Kurchatov Institute, Moscow 123182, 
Russia}
\mbox{$^6$ Department of Physics, Gustavus Adolphus College, St. Peter, MN 
56082, USA}
\mbox{$^7$ Department of Physics and Lawrence Berkeley National Laboratory, 
Berkeley, CA 94720, USA} 
\mbox{$^8$  Department of Physics, Jeonbuk National University, Jeonju City 
560-756, S. Korea}
\mbox{$^{9}$ Institute of Physics, University of Fribourg, CH-1700 Fribourg,
Switzerland} \\
\mbox{$^{10}$ Department of Physics, University of Wyoming, Laramie, WY 
82071-3905, USA} \\
\mbox{$^{11}$ Paul Scherrer Institute, CH-5232 Villigen, Switzerland} \\
\mbox{$^{12}$ TRIUMF, Vancouver, BC, Canada V6T 2A3} \\
\mbox{$^{13}$ Institute for Medium Energy Physics, Austrian Academy of 
Sciences, A-1090 Vienna, Austria}
}

\date{\today}
\maketitle
\begin{abstract}

Measurements of muon catalyzed $dt$ fusion ($d\mu t \rightarrow ^4\!\!\!He + n + 
\mu^-$)
in solid HD have been performed. The theory describing the energy
dependent resonant molecular formation rate for the reaction $\mu t + HD 
\rightarrow [(d\mu t)pee]^*$ is compared to experimental results in a 
pure solid HD 
target. Constraints on the rates are inferred through the
use of a Monte Carlo model developed specifically for the experiment. From the
time-of-flight analysis of fusion events in 16 and 37 $\mu$g$\cdot$cm$^{-2}$ 
targets, an average formation rate consistent with 
0.897$\pm(0.046)_{stat}\pm(0.166)_{syst}$ times the theoretical prediction was 
obtained.
\end{abstract}

% Positronium, muonium, muonic molecules and atoms
% few body systems
% fusion reactions

The idea of introducing a negative muon into a mixture of hydrogen isotopes to
catalyze fusion has been known for over 50 years \cite{ponomarev}. The muon 
will be quickly captured
by an atom, replacing the orbital electron. This muonic atom can then 
interact with an isotopic hydrogen molecule to form a muonic 
molecular ion. The negative muon is about 207
times the mass of an electron, thus the two nuclei in the molecule are bound 
together very tightly. The negative muon shields
the repulsive Coulomb force between the two nuclei and because of their close
proximity, fusion occurs and the muon can take part in another fusion cycle. 
The rate-limiting step in this process is muonic molecular (or molecular ion) 
formation. However, the probability of the muon sticking to the alpha particle
after fusion occurs limits the average number of fusions to approximately 
200 per muon.  

The theory describing the energy dependent resonant molecular formation rate 
for the reaction $\mu t + HD \rightarrow [(d\mu t)pee]^*$ has 
been developed in the last decade. The high rate is expected \cite{petitjean}
to enhance muon catalyzed fusion in triple mixtures of H/D/T. In 
this letter we report the results of a muon catalyzed $dt$ fusion experiment 
which confirm this rate. We have performed the first measurement using $\mu t$ 
atoms incident on pure solid HD, measuring energy via a time of flight 
technique. By using a Monte Carlo simulation (which 
employs the theoretical energy-dependent rates) we have compared the 
experimental results to the theory, thus determining the resonant molecular 
formation rate consistent with the fusion time spectra for two different HD 
target thicknesses.
        
\begin{figure}
\begin{center} \mbox{
\epsfig{file=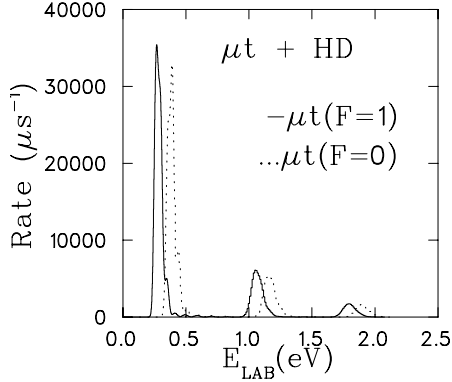,height=5.0cm}}
\end{center}
\caption{\label{resonances} Theoretically predicted molecular formation rate
calculated for 3K, normalized to liquid hydrogen density, for the two hyperfine
states of $\mu t$.}
\end{figure}

Figure~\ref{resonances} shows the theoretically predicted resonant molecular 
formation rate as a function of $\mu t$ lab energy. This calculation, 
normalized to liquid hydrogen atomic density (LHD), was performed for 
isolated HD molecules at a temperature of 3K as had been done at higher
temperatures \cite{faifman}. The rate for HD is then predicted by 
multiplying by the fractional deuteron concentration, $C_d$=0.5. The excess 
collisional energy of the system 
along with the released binding energy of the $d\mu t$ 
molecular ion is absorbed by excitation of the entire six-body complex 
$[(d\mu t)pee]$. Because the $d\mu t$ ion exists in a weakly bound state,  
resonance formation is possible without dissociation.

The experiment described here has the advantage of using
spatially separated layers of different isotopic composition 
in which interactions of the muonic system can be 
separated, providing the opportunity to measure the rate and energy dependence
of resonant molecular formation using time of flight. It is similar to our 
previous reported measurement \cite{Fujiwara} using D$_2$ molecules. 
Other experimental efforts in liquid/gaseous triple mixtures attempt to measure
this rate, however interpretation is difficult because of the complex kinetics
involved [5,6]. In the present experiment, solid layers were made by freezing 
hydrogen isotopic gases and mixtures onto two 51 $\mu$m 
thick gold target foils maintained at 3K. The gas mixtures
were deposited onto the foils through a
gas diffuser [7,8], where they could remain frozen for up to one week. 
For the HD target layer, commercial 96\% 
HD gas was passed through a molecular sieve
placed in a liquid nitrogen bath to remove nonhydrogen impurities, and was
subsequently monitored during target deposition with a quadrupole mass 
spectrometer to confirm nonequilibration of the gas. It has been
shown that at cryogenic temperatures the molecules do not begin to 
equilibrate for roughly 200 hours \cite{souers}. A nonhydrogen contamination 
following target deposition was determined to be less than 0.1\% through use 
of a germanium detector which measured x-rays from muon transfer to high-Z 
nuclei as well as gamma rays from nuclear capture. 

\begin{figure}
\begin{center} \mbox{
\epsfig{file=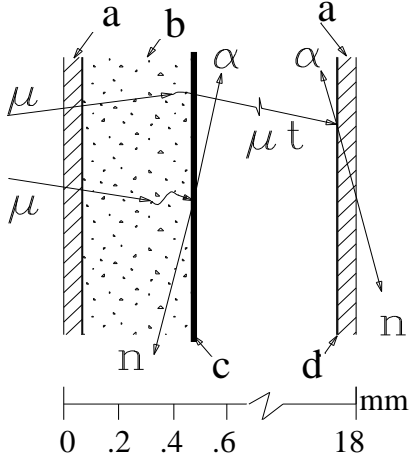,height=6cm}}
\end{center}
\caption{\label{foils} Solid hydrogen layer structure. Fusion can occur in 
either the D$_2$ 
moderation layer (c), or HD reaction layer (d).}
\end{figure}

Figure~\ref{foils} depicts the target and the muonic
interactions which occur. Negative muons having a momentum near 27 
MeV/c travel through one of the gold foils (a) into the 3.5 mg$\cdot$cm$^{-2}$ 
production 
layer (b) which consists of protium ($^1$H$_2$) containing 0.1\% tritium 
(T$_2$). Here the muon is atomically captured by a proton within a time of
$10^{-11}$ s (at LHD) and subsequently 
transfers to a triton due to the
increased binding energy of $\mu t$. The neutral $\mu t$ atoms, created with 
tens of eV energy, escape the production layer
before thermalization due to the Ramsauer-Townsend mechanism which reduces
their scattering cross section on H$_2$. The moderation
layer (c), consisting of 70 $\mu$g$\cdot$cm$^{-2}$ D$_2$, reduces the 
energy of the $\mu t$ atoms to optimize resonant molecular formation 
on the downstream HD target (d) of thickness 16 $\mu$g$\cdot$cm$^{-2}$. The
$\mu t$ atoms traverse 17.9 $\pm$ 0.1 mm in vacuum to reach (d). 
For those at or near the resonant energy, molecular 
formation and fusion follow with high probability, producing a 3.5 MeV alpha 
particle and a 14 MeV neutron. The $\mu t$ atoms which do not escape
the moderation layer (c) can interact there with deuterium, and subsequent 
$dt$ fusion can occur \cite{Fujiwara}.  

\begin{figure}
\begin{center} \mbox{
\epsfig{file=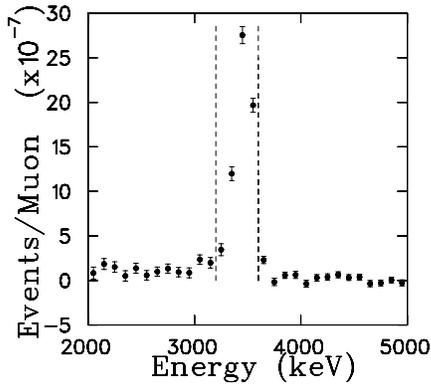,height=5.0cm}}
\end{center}
\caption{\label{energy} Energy deposited in silicon detectors by 
3.5 MeV fusion alpha particles from the reaction layer, normalized per muon 
($N_\mu$), for the 16 $\mu$g$\cdot$cm$^{-2}$ HD data. 
The dashed lines show the energy cut imposed for the time-of-flight analysis.}
\end{figure}

Alpha particles from fusion in (c) and (d) are separated by time of flight. 
The muonic atom emission time 
from the production layer, dominated by muon transfer from proton to triton, 
is of order 0.1 $\mu$s. Resonant formation 
rates are of order 10$^9$ s$^{-1}$, while $dt$ fusion occurs in the molecular 
ion at a rate of 
10$^{12}$ s$^{-1}$. Since the two gold foils are separated by 17.9 mm, a
$\mu t$ atom which travels through the moderator layer reaches the 
downstream HD layer in a minimum time of 0.2 $\mu$s, and more typically
2 $\mu$s at 1 eV. Thus, the time interval
between detection of a muon entering the target and an alpha from fusion
in the downstream HD reaction layer is dominated by the flight
time of the $\mu t$ atom. Two planar silicon 
detectors located perpendicular to the target foils, each with 
active area 2000 mm$^2$ and nominal depletion depth 300 $\mu$m, detect 
alpha particles. A
time cut of 1.5 to 6.0 $\mu$s, selects 
fusion events occurring in the HD layer only. The summed energy spectrum from
both silicon detectors, normalized to the number $N_\mu$ of good incident muons
(see \cite{Fujiwara} for details), is shown in Fig.~\ref{energy}.
Background was removed by subtracting normalized data for which no downstream
reaction layer was present.  

A Monte Carlo (MC) program was used to compare theoretical resonant molecular 
formation rates to data. This was necessary as the time-of-flight 
spectra cannot be uniquely inverted due to geometrical effects and the energy 
loss of 
$\mu t$ atoms in the HD reaction layer prior to molecular formation. 
The MC code \cite{huber} simulates muonic processes occurring in 
the experiment ({\it e.g.}, scattering, muon transfer, muonic molecular 
formation, fusion), taking into account the dimensions and geometry of the 
apparatus. In particular, it uses the rate of 
Fig.~\ref{resonances}, scaled by $C_d$=0.5, to describe resonant formation in 
HD. The majority of cross sections and rates used in the code were taken from 
\cite{bracci} and \cite{faifman2}. However, experimentally 
measured rates for muon transfer ($\mu p \rightarrow \mu t$) and nonresonant 
$p\mu p$ molecular formation \cite{mulhauser} were used (see also 
\cite{porcelli}). Values for the cross section $\sigma_{\mu t+d}$ equal to 0.9 
times those of \cite{bracci} were found to better represent the fusion time
distribution observed
in the moderation layer \cite{marshall}, and thus were used in the simulation.
These define our nominal MC input; the
code has been validated through 
comparison with independently written simulation code for simplified cases for 
which a direct comparison can be made \cite{markushin}.
 
To compare directly the time-of-flight fusion results to the simulation, 
the simulations were scaled by the silicon detector solid angle, 
$\Omega_{Si}=(2.32\pm 0.10)\times 10^{-2}$ and normalized to the fraction of 
incoming muons which 
stopped in the production layer, $S_F=(32\pm2)$\% \cite{marshall}.

\begin{figure}
\begin{minipage}{0.46\linewidth}
\centerline{\epsfig{file=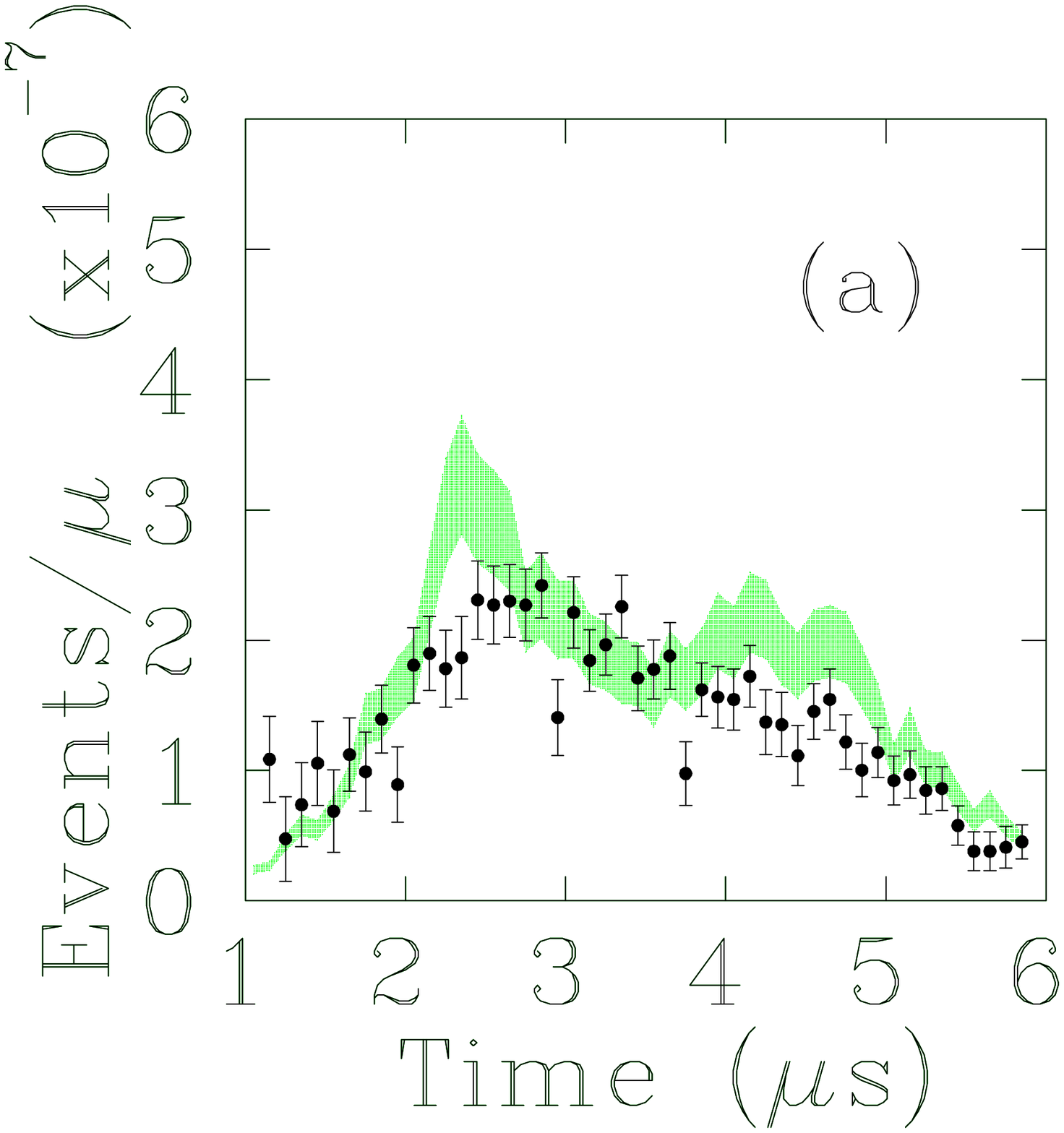,height=4.9cm}}
\end{minipage}
\hfill
\begin{minipage}{0.49\linewidth}
\centerline{\epsfig{file=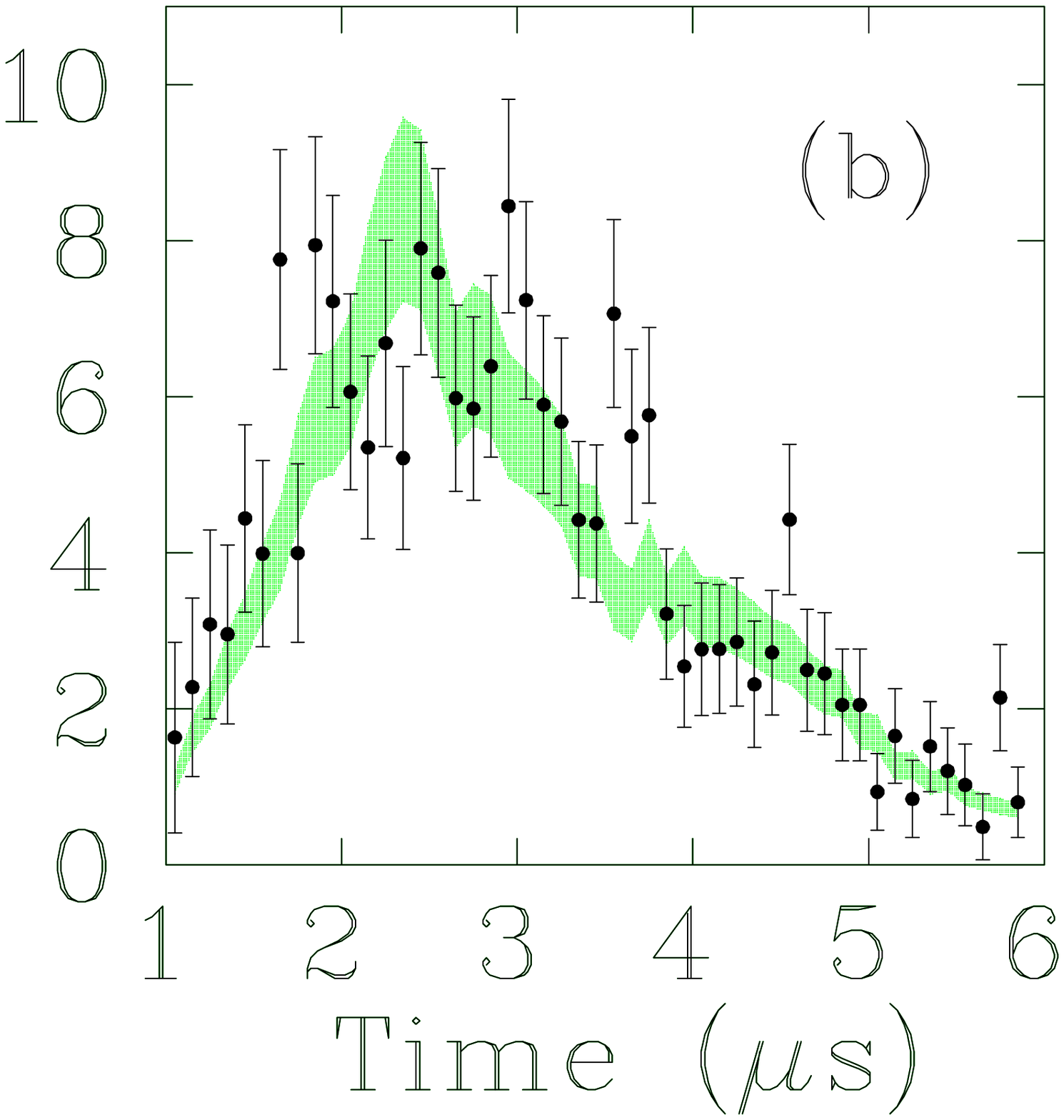,height=4.0cm}}
\end{minipage}
\caption{\label{results} Time-of-flight spectra for (a) 16 and (b) 37
$\mu$g$\cdot$cm$^{-2}$
HD targets. Data are plotted as solid dots with statistical error bars and 
simulations are shown as histograms with the width of the band indicative of 
the known systematic errors of the experiment.}
\end{figure}

The fusion time-of-flight data for two HD targets are shown in 
Fig.~\ref{results}. Both the simulation and data
are plotted as fusion events per incident muon $N_\mu$. 
Figure~\ref{results} (a) shows the 
results for the
16 $\mu$g$\cdot$cm$^{-2}$ HD target. The simulation shows a clear 
two-peaked 
structure not evident in the data. One peak between 2 and 3
$\mu$s corresponds to the molecular formation resonances which occur 
above 1.0 eV, while the other,  between 4 and 5 $\mu$s, corresponds to the
stronger resonances near 0.3 eV. Data from the 37 $\mu$g$\cdot$cm$^{-2}$ HD 
reaction layer are shown in Fig.~\ref{results} (b). 
Here, the $\mu t$ atoms will undergo more scattering reactions before
molecular formation occurs and hence are less sensitive to the
molecular formation resonances. This is clearly visible in the simulation 
results, where the two-peaked structure is less evident. The agreement
between simulation and data for this target thickness is quite reasonable (see
below). Good agreement was obtained also for the case of a D$_2$ target layer,
as reported previously \cite{Fujiwara}.

To test sensitivity of our data to resonant molecular formation rates, several
modifications to the nominal MC input were made. Differences from the nominal
input (a) include the following: determining the
effect of a decreased $\sigma_{\mu t+d}$ scattering cross section (b); shifting
the positions of the resonances to higher (c) and lower (d) energies; using an
energy independent constant rate to model molecular formation (e); and 
broadening the molecular formation resonances (f). Table~\ref{chart} summarizes
the target thickness, input changes to the MC, the fit ratio $R_F$ which is the 
factor by which the data must be scaled to provide the best fit to the 
simulation, and the $\chi^2$ per degree of freedom (dof=49) for the scaled
fit. To provide a check of how well the simulation is reproducing the absolute
intensity observed, the $\chi^2$/dof of fits are listed for $R_F$ being held 
fixed at 1. 

To investigate sensitivity to 
the positions of the molecular formation resonances, they were
shifted in the simulation by 0.18 eV (the average standard deviation of the 
peaks) to higher 
and lower energies. The peaks appearing in
the simulated fusion time distribution shifted as expected due to the
inverse relationship between the time of flight and speed of the $\mu t$ atom.
The results given in Table~\ref{chart} show increases
in $\chi^2$/dof for both the shape and intensity tests 
when compared to the simulation results using the
nominal predicted positions of the resonances. The results are indeed 
sensitive to the position of the resonances, but shifting them does not 
improve agreement between simulation and data.

A simulation using a constant molecular formation rate of $\lambda_{d\mu t}=
200$ $\mu s^{-1}$, chosen to approximately match the intensity of the data, 
was also compared to data. The $\chi^2$/dof for the fit to the 16 $\mu g\cdot$
cm$^{-2}$ data is 2.99 compared to 1.87 using the nominal MC input. 
 
To determine whether the discrepancy in shape could be 
explained by resonance broadening, the theoretical resonances were convoluted
with a Gaussian of 50 meV rms, while preserving the area under each resonance.
This MC time distribution for the 37 $\mu$g$\cdot$cm$^{-2}$ data 
did not reproduce the data as well as the nominal simulation.
For the 16  $\mu$g$\cdot$cm$^{-2}$ data, the two-peaked structure remained in 
the MC, with the distribution of both peaks quite broad and 
a factor of 1.6 higher in yield than the data for the time range between 3.0
and 6.0 $\mu s$. 

To determine the best value for the number ($S_\lambda$) by which the resonant 
molecular formation rate should be scaled to agree with the data,
simulations were done for $S_\lambda$ between 0.50 and 1.40 ($S_\lambda=1.0$
corresponds to agreement with theory) and the results were then fit to the 
data. The $\chi^2$ values of these fits versus $S_\lambda$
allow estimation of the scaling factors for the 16 and 37 
$\mu$g$\cdot$cm$^{-2}$ targets of, respectively,  
0.798 $\pm$ 0.060 and 1.040 $\pm$ 0.072, which are not in agreement within 
their quoted statistical errors. As
the significant known systematic errors in the experiment (such as solid angle
and stopping fraction) are common to both the 16 
and 37 $\mu$g$\cdot$cm$^{-2}$ targets, they cannot be used to reconcile the 
scaling factors, pointing to an
unknown systematic error or inadequacy of the simulation for which an 
additional systematic uncertainty of unknown origin must be determined. Using 
the
method described in \cite{pp}, where the assessed uncertainty is increased 
to reflect possible unknown sources of error, an uncorrelated systematic 
uncertainty of 2.38 times the statistical uncertainty is added in quadrature
with the correlated systematic uncertainty. The 
average scaling factor for both targets is then found to be 0.897 
$\pm (0.046)_{stat} \pm (0.166)_{syst}$.

The scaling factors for the resonant molecular formation rates for the targets 
studied are given in Table~\ref{final}. The experimental results confirm the
high HD resonant rates expected at these energies. However, the fusion time
spectra structure is not confirmed, in contrast to the results for D$_2$
\cite{Fujiwara}. The two-peaked structure predicted by 
the Monte Carlo results for the thin HD target is not apparent in the data. 
Several adaptations to the resonant molecular formation theory were 
investigated, 
but this peaked structure persists in the MC. On the other hand, using a 
constant rate to describe
molecular formation modeled the data less accurately than the
resonance model. In the thicker HD target where
resonant molecular formation competes with scattering of the $\mu t$ atoms
and hence energy loss, agreement between theory and data is quite good. The
theory developed for the resonant process does not take into account any 
effects of the crystal lattice or its motion with respect to the $\mu t$ 
atoms. It has been suggested that effects in the solid hydrogen crystal lattice
may affect the $\mu t$ slowing process and thus the reaction yields 
\cite{adamczak}. 
Incorporating such effects into the simulation is a logical next step.

This work was supported by NSERC (Canada), the DOE and NSF (USA), the Swiss 
National Science Foundation and a NATO Linkage Grant LG 930162.

\begin{table}[h]
\begin{tabular}{lcccc}
&MC& $R_F$&$\chi^2$/dof&$\chi^2$/dof \\
&Input& & & ($R_F$=1) \\ \hline
\multicolumn{4}{c} {HD layer = 16 $\mu$g$\cdot$cm$^{-2}$}\\ \hline
(a) &nominal (see text)&0.787(29)&1.87&3.23 \\   
(b) &$\sigma_{\mu t +d}$ unscaled& 0.839(31) & 1.93 & 2.77 \\  
(c) &+0.18 eV shift & 0.753(52)& 5.86& 7.84\\  
(d) &-0.18 eV shift& 0.952(51) &3.69& 3.69 \\ 
(e) &$\lambda_{d\mu t}$=200 $\mu s^{-1}$ &1.03(5)&2.99&2.99\\ 
(f) &convolution&0.639(21)&1.63&9.54\\ \hline
\multicolumn{4}{c} {HD layer = 37 $\mu$g$\cdot$cm$^{-2}$} \\ \hline
(a)& nominal (see text)&1.04(4)&1.27&1.29 \\ 
(b) & $\sigma_{\mu t + d}$ unscaled & 1.14(5)& 1.27& 1.47 \\ 
(c) & +0.18 eV shift & 1.06(6)& 2.32&2.32\\ 
(d) & -0.18 eV shift& 1.08(5)&1.40&1.46\\ 
(e) & $\lambda_{d\mu t}$=200 $\mu s^{-1}$ & 0.998(37)&0.97&0.97\\ 
(f) & convolution&0.751(29) & 1.11 &2.69  \\
\end{tabular}
\caption{\label{chart} Listed are the changes in Monte Carlo input, the fit 
ratio $R_F$
along with the $\chi^2$/dof for the fit. Fits were
also done holding the scaling of data to MC fixed at $R_F$=1.00, with the 
resulting $\chi^2$/dof given.}
\end{table} 

\begin{table}[h]
\begin{tabular}{cc}
Target& $\lambda_{d \mu t}$ Scaling Factor ($S_\lambda$) \\ \hline
16 $\mu$g$\cdot$cm$^{-2}$&0.798 $\pm$ (0.060)$_{stat}$ $\pm$ (0.110)$_{corr}$
 \\
37 $\mu$g$\cdot$cm$^{-2}$&1.040 $\pm$ (0.072)$_{stat}$ $\pm$ (0.148)$_{corr}$
\\
weighted average & 0.897 $\pm$ (0.046)$_{stat}$ $\pm$ (0.166)$_{syst}$
\end{tabular}
\caption{\label{final} The values of $\lambda_{d\mu t}$ scaling factors with 
all sources of error included ($stat$=statistical, $corr$=correlated systematic
and $syst$=total systematic error). A value equal to 1.0 is predicted 
theoretically.}
\end{table}


\begin{references}
\bibitem[*]{newadd} Present address: Department of Physics, University of 
Northern British Columbia, Prince George, B.C., Canada V2N 4Z9; 
email: Porcelli@triumf.ca.
\bibitem[\dagger]{newadd2} Present address: Department of Physics, University 
of Tokyo, Hongo, Tokyo 113-0033, Japan. 
\bibitem{ponomarev}L.I. Ponomarev, Cont. Phys. {\bf 31}, 219 (1990).
\bibitem{petitjean}C. Petitjean, Nucl. Phys. A {\bf 543}, 79c (1992).
\bibitem{faifman}M.P. Faifman {\it et al.}, Hyp. Int. {\bf 101-102}, 179
(1996); M.P. Faifman and L. Ponomarev, Phys. Lett. B {\bf 265}, 201 (1991).
\bibitem{Fujiwara}M.C. Fujiwara {\it et al.}, Phys. Rev. Lett. {\bf 85}, 1642
(2000).
\bibitem{jeitler}M. Jeitler {\it et al.}, Phys. Rev. A {\bf 51}, 2881 (1995).
\bibitem{yup}Yu.P. Averin {\it et al.}, Hyp. Int. {\bf 118}, 121 (1999).
\bibitem{knowles}P.E. Knowles {\it et al.}, Nucl. Instrum. Methods A 
{\bf 368}, 604 (1996).
\bibitem{porcelli}T.A. Porcelli, Ph.D. Thesis, University of Victoria, (1999). 
\bibitem{souers}P.C. Souers, Hydrogen Properties for Fusion Energy, University
of California Press, Berkeley, 1986. 
\bibitem{huber}T.M. Huber {\it et al.}, Hyp. Int. {\bf 118}, 159 (1999). 
\bibitem{bracci}L. Bracci {\it et al.}, Muon Catal. Fusion {\bf 4}, 247 (1989);
C. Chiccoli {\it et al.}, {\it ibid} {\bf 7}, 87 (1992).
\bibitem{faifman2}M.P. Faifman, Muon Catal. Fusion {\bf 4}, 341 (1989). 
\bibitem{mulhauser}F. Mulhauser {\it et al.}, Phys. Rev. A {\bf 53} 3069 (1996).
\bibitem{marshall}G.M. Marshall {\it et al.}, Hyp. Int. {\bf 118}, 89 (1999).
\bibitem{markushin}V.E. Markushin, Hyp. Int. {\bf 101-102} 155 (1996).
\bibitem{pp} Particle Data Group, Eur. Phys. J. C {\bf 15}, 11 (2000).
\bibitem{adamczak} A. Adamczak, Hyp. Int. {\bf 119}, 23 (1999).
\end{references}
\end{document}